\documentclass[twocolumn,amssymb,nobibnotes, aps, pra]{revtex4-2}

\usepackage{xcolor}
\usepackage{physics}
\usepackage{amsmath, amssymb}
\usepackage{graphicx}
\usepackage[colorlinks, linkcolor=blue, citecolor=blue, urlcolor=blue, breaklinks=true]{hyperref}
\usepackage{mathtools}
\usepackage{float}

\begin{document}
\title{Time optimal quantum state engineering }

\begin{abstract}
The efficient generation of highly nonclassical quantum states is essential for emerging quantum technologies, yet it  remains challenging due to decoherence and the long preparation times associated with conventional adiabatic protocols. Here, we employ time-optimal control methods based on the quantum brachistochrone formalism to engineer nonclassical states in light--matter systems described by the time-dependent Jaynes--Cummings and quantum Rabi models. We demonstrate the deterministic generation of Fock states and highly entangled Schr\"odinger cat states at unit fidelity, and characterize their nonclassical properties through joint Wigner phase-space distributions. We further show that the wind control generates these non-classical states at speed limit which 
leads to a reduced energetic cost and robustness against dissipation, relaxation, and dephasing across a broad range of environmental conditions. Our results establish time-optimal control as an efficient and experimentally feasible approach for fast and nonclassical state engineering in hybrid quantum platforms. 
\end{abstract}

\author{Sam Edmunds}
    \email{S.Edmunds2@newcastle.ac.uk}
\author{Obinna Abah} 
\email{obinna.abah@ncl.ac.uk}
    \affiliation{School of Mathematics, Statistics and Physics, Newcastle University, Newcastle upon Tyne NE1 7RU, United Kingdom}

\date{\today} 


\maketitle

\section{Introduction} \label{sec:intro}

Many emerging quantum technologies rely on the ability to generate target quantum states with high fidelity, since imperfections in state preparation directly degrade the performance of quantum devices and protocols \cite{Bassi2013}. 
Consequently, the design of efficient protocols capable of generating target states with near-unit fidelity remains a central objective of quantum control and quantum information science \cite{Koch2022}.
Quantum resources, such as Schrödinger cat states represent paradigmatic examples of highly nonclassical states, consist of coherent superpositions of macroscopically distinguishable configurations. In light-matter systems \cite{larson2021jaynes,Jiang2025}, these states are read as $\ket{\mathrm{Cat}_{\pm}}\!=\!\left(
\ket{\alpha}
\pm
e^{i\theta}
\ket{-\alpha}
\right)/\mathcal{N}_{\pm}$,
where $\ket{\pm\alpha}$ denote coherent states,
$\mathcal{N}_{\pm}\!=\!\sqrt{2\left(1\pm e^{-2|\alpha|^2}\right)}
$
is the normalization factor, and $\theta$ determines the relative phase between the coherent components. The generation of entangled cat states and their nonclassical interference properties, cat states, has attracted significant interest for applications in quantum information processing \cite{gilchrist2004schrodinger,li2017cat, mirrahimi2014dynamically}, quantum metrology \cite{joo2011quantum}, quantum spectroscopy\cite{kira2011quantum,Wright2020}, and quantum error correction \cite{schlegel2022quantum}. In addition, bosonic cat-state encodings have emerged as promising candidates for fault-tolerant quantum computing due to their resilience against decoherence channels \cite{li2017cat}, and their ability to enhance qubit lifetimes \cite{schlegel2022quantum}.

The generation of nonclassical states, such as Fock states \cite{PhysRevLett.88.143601, hofheinz2008generation, uria2020deterministic, zhang2024generating}, entangled states \cite{kumar2023experimental, cao2023generation, munoz2023phase, santos2022generating, bartolucci2021creation, li2023speeding} and Schr\"odinger cat states \cite{cosacchi2021schrodinger, takase2021generation,kudra2022robust,in2021generating}, has been extensively investigated in a variety of physical systems, including cavity and circuit quantum electrodynamics platforms \cite{grimm2020stabilization,Hoshi2025}, and trapped ions \cite{Monroe1996,Poschinger2010,Johnson2017}, among others. For example, a light-matter coupled system in a deep-strong coupling regime can generate an entangled cat state, and the effective coupling strength is adiabatically controlled, e.g., a time-dependent parametric drive \cite{Leroux2018}. However, to avoid long evolution times, some preparation approaches include coherent driving protocols \cite{warren1993coherent, morzhin2023optimal, jiang2024coherent, mena2024room, vaneecloo2022intracavity}, measurement-based methods \cite{PhysRevResearch.6.023159, volya2024state, rivera2024quantum, puente2309quantum}, dissipative state preparation \cite{lambert2310fixing, mi2024stable}, and recently shortcut-to-adiabaticity (STA) techniques \cite{abah2020quantum,innocenti2020ultrafast,chen2021shortcuts,xu2024optimally, Yu2025, chen2025experimental} have been explored. Although adiabatic and STA-based protocols can achieve high-fidelity state preparation, conventional adiabatic methods require long evolution times that increase susceptibility to decoherence and operational errors and STA methods are often designed to follow specific adiabatic trajectories and may involve experimentally challenging control fields \cite{abah2020quantum,chen2021shortcuts}.

Quantum optimal control provides an alternative strategy by  optimizing the system dynamics such that system system is driven towards a desired target state while satisfying experimental constraints \cite{Glaser2015,Koch2022}. Rather than enforcing adiabatic evolution, optimal control seeks control fields that maximize target-state fidelity while minimizing physical resources such as energy or preparation time. This approach has become a powerful tool for steering quantum dynamics and for generating complex quantum states that are not accessible through conventional protocols. Studies have further demonstrated the potential of optimized  optimal-control functionals tailored for preparing arbitrary and entangled cat states while identifying associated quantum speed limits \cite{Glaser2015}.
However, understanding of the energetic cost \cite{Li2022} or quantum-energy performance \cite{abah2019energetic,Auffeves2022} has  become crucial for the practical implementation of  quantum technologies. 
Recently, it has been shown that the time-optimal control via quantum wind control provides a general framework for transforming  arbitrary pure states within a Hilbert space in the presence of the background "wind" (i.e the uncontrolled background Hamiltonian) while minimizing the evolution time under  energetic cost co \cite{garcia2022highly}. It has been experimentally demonstrated that this protocol provide energy-efficient quantum control while maintaining high fidelity, achieving  highly time-optimal adiabatic quantum driving with low energy cost throughout the whole evolution \cite{Xu_2024, Dong2024, Dong2025}. 


In this paper, we employ time-optimal control methods \cite{PhysRevLett.96.060503, PhysRevA.75.042308, carlini2008time} based on the quantum brachistochrone problem, known as the "quantum wind" \cite{garcia2022highly}, to engineer highly nonclassical entangled Schrödinger cat states in light-matter systems described by the quantum Rabi model. This model describes the interaction between a two-level system and a quantized bosonic mode, in the weak- and strong-coupling regimes, and provides experimentally relevant platforms for realizing hybrid quantum states. By minimizing the evolution time required to reach the target state, this approach enables us to prepare entangled cat states with unit fidelity while simultaneously reducing exposure to decoherence. This demonstrates that time-optimal quantum control provides an efficient and robust route for generating highly non-classical states in realistic quantum architectures.


The remainder of this paper is organised as follows. We begin in Sec.~\ref{section2} by briefly introducing the time optimal control framework. In Sec. \ref{sec:Light matter interaction}, we present the physical model of the system, Rabi Hamiltonian, and demonstrate the possibility of generating nonclassical states in different coupling regimes.  The nonclaissicality of the generated state is characterized using Wigner phase space function in Sec. \ref{nonclassicality}, and the protocol performance in Sec. \ref{sec.cost}.  
In Sec. \ref{sec:Noise}, we examine the robustness of time optimal protocol against the decay and decoherences of the environment. 
   Finally, the discussions and conclusions are presented in Sec. \ref{sec:Conculsion}.




\section{Time-optimal control protocol}\label{section2}
\begin{figure}[t] 
    \includegraphics[width = 0.5\textwidth]{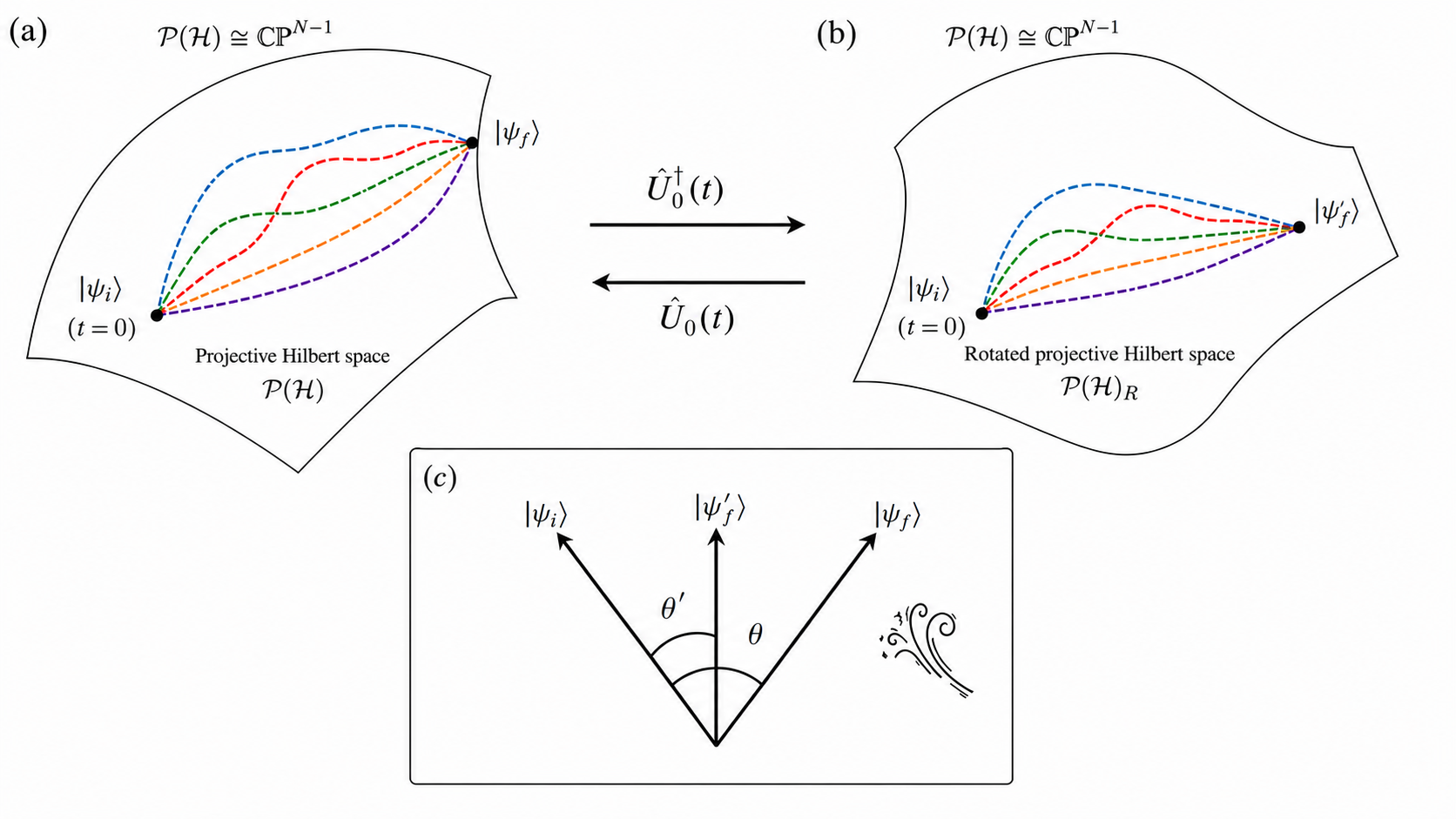}
    \caption{Schematic of the possible trajectories of the quantum state $\ket{\psi(t)}$ through the ray space/projected space $\mathcal{P}(\mathcal{H}) \cong \{\, \ket{\psi} \in \mathcal{H} : \bra{\psi}\ket{\psi} = 1 \,\} / U(1)$ in the (a) Schr\"{o}dinger picture and (b)interaction picture. Moving between these two frames via the rotation $\hat{U}_0(t)$ can lead to a "favourable wind" i.e the length of the paths connecting the initial $\ket{\psi_i}$ and $\ket{\psi_f}$ may decrease. This is depicted in (c) where a favourable wind decrease the Bures angle $\theta = 2\arccos|\braket*{\psi_{i}}{\psi_{f}}|$ to $\theta' = 2\arccos|\braket*{\psi_{i}}{\psi'_{f}}|$, where $\ket{\psi'_{f}} = \hat{U}^\dagger_0(t)\ket{\psi_{f}}$. }
\label{fig:manifold pic}
\end{figure}

State preparation can be described as the mapping between two normalized states in a $N$-dimensional Hilbert space $\ket{\psi_i}, \ket{\psi_f} \in \mathcal{H}$. Such a transformation can be realized through a single unitary operator $\hat{\mathcal{U}}$, according to $ \ket{\psi_f} = \hat{\mathcal{U}}\ket{\psi_i}$,
where $\ket{\psi_i}$ and $\ket{\psi_f}$ denote the initial and target states, respectively.  In general, this mapping is not unique since there exists a family of unitary operators satisfying the transformation;
\begin{equation}
    \mathfrak{S}=\left\{
\hat{\mathcal{U}} \in \mathcal{U}(\mathcal{H})\;\middle|\;
\hat{\mathcal{U}}\ket{\psi_i}=\ket{\psi_f}
\right\}.
\end{equation}
To generate any unitary map, $\hat{\mathcal{U}} \in \mathfrak{S}$, the state  must evolve under a Hamiltonian $\hat{H}(t)$ such that 
\begin{equation} \label{U generation}
\hat{\mathcal{U}} = \hat{U}(\tau), \quad \hat{U}(t) = \mathcal{T} \exp\left(-i \int_0^{t} \hat{H}(t_1)\, dt_1 \right),
\end{equation} 
where $ \hat{U}(t)$ is the time evolution operator, $\mathcal{T}$ is the time ordering operator, and $\tau$ is the total time the state evolves $\ket{\psi(t)} = \hat{U}(t)\ket{\psi_i}$ under $\hat{H}(t)$. Since Eq.~(\ref{U generation}) is also non-unique, different Hamiltonians $\hat{H}(t)$ can generate the same unitary $\hat{\mathcal{U}} = \hat{U}(\tau)$, generally for different evolution times $\tau$. Therefore, the state evolution may follow different trajectories $\ket{\psi(t)}$ through the projective Hilbert space $\mathcal{P}(\mathcal{H})$ depending on the choice of $\hat{H}(t)$, as illustrated in fig.~\ref{fig:manifold pic}(a). Now, consider a given initial state $\ket{\psi_i}$ evolving under a time-dependent Hamiltonian $\hat{H}(t)=\hat{H}_0(t)$. To find the time-optimal path (geodesic), i.e, the trajectory through $\mathcal{P}(\mathcal{H})$ that implements the target unitary $\hat{\mathcal{U}}\!=\!\hat{U}(\tau)$ in minimal time $\tau$ one can 
implement a control Hamiltonian $\hat{H}_c(t)$, such that the system dynamics is governed by $\hat{H}(t) = \hat{H}_0(t) + \hat{H}_c(t)$ \cite{garcia2022highly}. 
In the interaction picture, denoting the initial and final states, respectively, as $\ket*{\psi'_i}=\ket{\psi_{i}}$, $\ket*{\psi'_f}=\hat{U}^\dagger_0(\tau)\ket{\psi_{f}}$, the state $\ket*{\psi'(t)} = \hat{U}_{0}^{\dagger}(t)\ket*{\psi(t)}$ evolves along the geodesic according to $i \partial_{t}\ket*{\psi'(t)}=\hat{H}'_{c}(t)\ket*{\psi'(t)}$ where $\hat{U}_{0}(t)=e^{-i\int_{0}^{t}\hat{H}_{0}(t_{1})dt_{1}}$ and 
\begin{equation} \label{control Hamiltonain}
    \hat{H}'_{c}(t) = i \frac{v_{z}(t)}{\sqrt{1-s^2}} \left( e^{-i \beta}\ket*{\psi_{f}'}\bra{\psi_{i}}-h.c \right)
\end{equation}
with $se^{i \beta} = \braket*{\psi_{i}}{\psi'_{f}}$ being the overlap between the initial state and the final state in the interaction picture and $v_{z} = \Delta\hat{H}'_{c}(t)$ being the variance of the control Hamiltonian where $\Delta X^2\!=\! 
\langle \psi(t) | X^2 | \psi(t) \rangle
- \left(\langle \psi(t) | X | \psi(t) \rangle\right)^2$.
The state evolving along this minimum path between the initial and the final states reads 
\begin{equation} \label{geodesic path}
    \ket*{\psi'(t)} =  \left (\cos\theta(t)-\frac{\sin{\theta(t)}}{\sqrt{1-s^2}} \right)\ket{\psi_{i}}+\frac{e^{-i \beta}\sin{\theta(t)}}{\sqrt{1-s^2}}\ket*{\psi'_{f}}
\end{equation}
where $\theta(t) = \int_{0}^{t}\Delta H_c'(t_1) dt_1$.
By definition, this path minimizes the Riemannian metric of Fubini and Study, which in the interaction picture is given by Eq.~\ref{int FS metric} 
\begin{equation} \label{int FS metric}
    d\mathcal{L}_{FS}^2 = 4\left(\frac{\bra{d\psi'}\ket{d\psi'}}{\bra{\psi'}\ket{\psi'}} -\frac{|\bra{\psi'}\ket{d\psi'}|^2}{\bra{\psi'}\ket{\psi'}^2}\right).
\end{equation}
By substituting Eq.~(\ref{geodesic path}) into Eq.~(\ref{int FS metric}),   the length of the path followed by a normalized quantum state in $\mathcal{P}(\mathcal{H})$ is
\begin{equation} \label{length of path}
    \mathcal{L}_{FS} = 2\int_{0}^{\tau} v_z(t) dt \geq \mathcal{L}(\ket{\psi_i},\ket*{\psi_f'}),
\end{equation} 
where for any two states $\ket{\psi_i},\ket{\psi_f} \in \mathcal{H}$
\begin{equation}
    \mathcal{L}(\ket{\psi_i},\ket*{\psi_f})\!=\!2\arccos\left(\sqrt{F(\ket{\psi_i},\ket*{\psi_f})}\,\right)
\end{equation}
is the Bures angle, i.e., the minimum distance between $\ket{\psi_i},\ket{\psi_f}$, $v_z=\Delta H(t)$ is the quantum "speed" and
\begin{equation}
    F(\ket{\psi_i},\ket{\psi_f}) =|\bra{\psi_i}\ket{\psi_f}|^2 
\end{equation}
is the fidelity. In order for the control to truly evolve the state along the geodesic it must 
satisfy the equality in Eq.~(\ref{length of path}). Assuming that the energy resource associated with the control is constant at maximum value, i.e, $v_{z}(t)\!=\!v_{z}$, the relation between the velocity and the protocol time is find as \cite{garcia2022highly}
\begin{equation} \label{wind speed limit}
    v_{z} = \frac{1}{\tau} \arccos{\sqrt{\mathcal{F}_{0}}}
\end{equation}
where $\mathcal{F}_{0} = |\braket*{\psi_{i}}{\psi'_{f}}|^2$.
We remark, since Eq.~\ref{length of path} is computed in the interaction picture, the length of the geodesic $\mathcal{L}(\ket{\psi_i},\ket*{\psi_f'})$ can be smaller/larger compared to the Schr\"{o}dinger picture representation, see Fig.~\ref{fig:manifold pic}(b) and Fig.~\ref{fig:manifold pic}(c). 

\section{model} \label{sec:Light matter interaction} 
The simplest and most fundamental light-matter system used to generate entangled cat states is the \textit{Rabi model} (RM) \cite{PhysRev.51.652}, which describes the interaction between an electromagnetic cavity (bosonic mode) and a two-level system (qubit). The Hamiltonian of this model reads (taking $\hbar=1$),
\begin{equation} \label{RM model0}
    \hat{H}_{RM}(t) = \omega_{c} \hat{a}^{\dagger}\hat{a} + \frac{\omega_{q}(t)}{2}\hat{\sigma}_{z} + \lambda(t)\hat{\sigma}_{x}(\hat{a}^{\dagger} + \hat{a})
\end{equation}
where $\omega_c$ ($\omega_q(t)$) is the bosonic mode (qubit) frequency, $\lambda(t)$ is the light-matter coupling strength, $\hat{a}^{\dagger}$ ($\hat{a}$) is the bosonic creation (annihilation) operator, and $\hat{\sigma}_{x} = \ket{e}\bra{g}+\ket{g}\bra{e}$ and $\hat{\sigma}_{z} = \ket{e}\bra{e} - \ket{g}\bra{g}$ are Pauli operators, with $\ket{e}$ ($\ket{g}$) denoting the excited (ground) state of the qubit. The Rabi model has several distinct coupling regimes characterized by the normalized coupling strength $\eta\!=\! \lambda/\omega_c$: the weak $\eta \ll1$, ultrastrong $\eta \sim 0.1-1$, deep strong $\eta \gtrsim1$, and extreme strong coupling regimes $\eta \gtrsim 10$, 
which can all be used to effectively generate non-classical states. 

\begin{figure}[t] 
\hspace*{-0.5cm}
    \includegraphics[width = 0.5\textwidth]{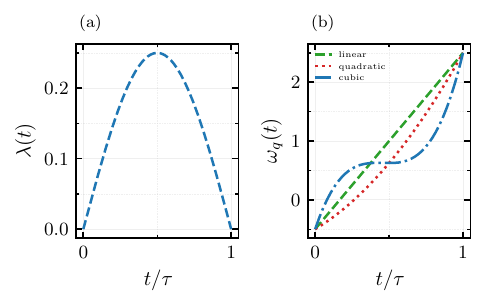}
    \caption{(a) The time-dependent coupling $\lambda(t)$,  Eq.~(\ref{coupling protocol}), as a function $t/\tau$  for $\lambda_0\!=\!0$, $\lambda_m\!=\!\omega_c/4$ and $\phi=0$. (b) The qubit driving frequency $\omega_q(t)$,  Eq.~(\ref{TLS driving}), as a function of $t/\tau$ for different driving profiles, see Eq.~(\ref{TLS profiles}). The parameter used are $\omega_i\!=\!-\omega_c/2$, $\omega_f=5\omega_c/2$, $\alpha=0.5$ and $\beta=2$. 
}
\label{fig:external driving}
\end{figure}

\emph{Weak-coupling regime} -- 
Assuming, the qubit and cavity are close to resonance, $\Delta\!=\!|\omega_q-\omega_c| \ll \omega_q+\omega_c$ and weakly coupled $\lambda \ll \omega_q,\omega_c$, the counter-rotating term 
$\lambda(\hat{a}\hat{\sigma}_{-}+\hat{a}^{\dagger}\hat{\sigma}_{+})$, can be neglected using the rotating-wave approximation (RWA) \cite{LarsonBook}.  The result of this approximation is the celebrated Jaynes-Cummings (JC) model \cite{1443594} $\hat{H}_{JC}(t) = \omega_{c} \hat{a}^{\dagger}\hat{a} + \frac{\omega_{q}(t)}{2}\hat{\sigma}_{z} + \lambda(t)(\hat{a}\hat{\sigma}_+ +\hat{a}^\dagger\hat{\sigma}_- )$ which is known to conserve the total number of excitations in the system, i.e., $[\hat{N}, \hat{H}]\!=\!0$ where $\hat{N}\!=\! \hat{a}^\dagger\hat{a} + \ket{e}\bra{e}$. Hence, the JC Hamiltonian can be block diagonalized in the excitation number subspace $\mathcal{H}_n\!=\!\{\ket{e,n}, \ket{g,n+1}\}$ where $n$ is the excitation number of the cavity  mode and $\mathcal{H} = \bigoplus_n \mathcal{H}_n$. \newline 
\newline 
Therefore, we can write $\hat{H}_{JC} = -\omega_q(t)\ket{g,0}\bra{g,0} + \oplus_n\hat{H}_n(t)$, where the Landau-Zener-like terms read as follows,
\begin{equation}
H_n =
\begin{pmatrix}
\omega_q(t)/2 + n\omega_c & \lambda(t)\sqrt{n+1} \\
\lambda(t)\sqrt{n+1} & -\omega_q(t)/2 + (n+1)\omega_c
\end{pmatrix}.
\end{equation}
We consider the following coupling and qubit driving protocols over time $t \in [0,\tau]$,
\begin{equation} \label{coupling protocol}
    \lambda(t) = (\lambda_m -\lambda_0)\sin(\mathcal{S}\pi + \phi) + \lambda_0
\end{equation}
where $\lambda_m$ is the maximum coupling value, $\lambda(0)\!=\! \lambda(\tau)\!=\!\lambda_0$, $\mathcal{S}\!=\!t /\tau$, $\phi$ is a  phase shift, and 
\begin{equation} \label{TLS driving}
    \omega_q(t) = \omega_i + (\omega_f - \omega_i)\, f(\mathcal{S}), \quad 0 \le f(\mathcal{S}) \le 1,
\end{equation}
with $\omega_q(0)=\omega_i$, and $\omega_q(\tau)=\omega_f$. In Fig.~\ref{fig:external driving}, we illustrate the behavior of the protocol $\lambda(t)$ and $\omega_q(t)$ for linear, quadratic, and cubic driving, 
\begin{equation} \label{TLS profiles}
f_j(\mathcal{S}) = 
\begin{cases}
\mathcal{S}, & j=1 \\
\mathcal{S} + \alpha\, \mathcal{S}(\mathcal{S}-1), & j=2 \\
\mathcal{S} + \alpha\,\mathcal{S}(\mathcal{S}-1) + \beta\, \mathcal{S}(\mathcal{S}-1)(2\mathcal{S}-1), & j=3
\end{cases}
\end{equation}
where $j\!=\!1,2,3$ corresponds to the linear, quadratic, and cubic driving of the qubit, respectively.

Using the qubit frequency and coupling introduced above, the JC model can be used to generate the $n^{th}$ Fock state  through a sequence of adiabatic population inversions, achieved by driving the qubit frequency $\omega_{q}(t)$ through the avoided energy gap while satisfying the adiabatic condition $\tau \gg 1$. To illustrate this process, 
consider the initial state given by $\ket{\psi_i}\!=\!\ket{e, 0}$, such that the system initially evolves entirely within the zeroth excitation subspace. Then $\omega_q(t)$ is slowly driven through the avoided crossing, i.e, adiabatic evolution, to ensure that the system remains in the same eigenstate. Consequently, the qubit and cavity populations are exchanged, leading to
$\ket{g, 1}$.  Afterward, a $\pi$-pulse is applied to excite the qubit,  obtaining the state $\ket{e,1}$ and thereby transferring  the system into the first excitation subspace. Repeating this sequence then gives $\ket{e,n}$.

\begin{figure}[t] 
\hspace*{-0.5cm}
    \includegraphics[width = 0.5\textwidth]{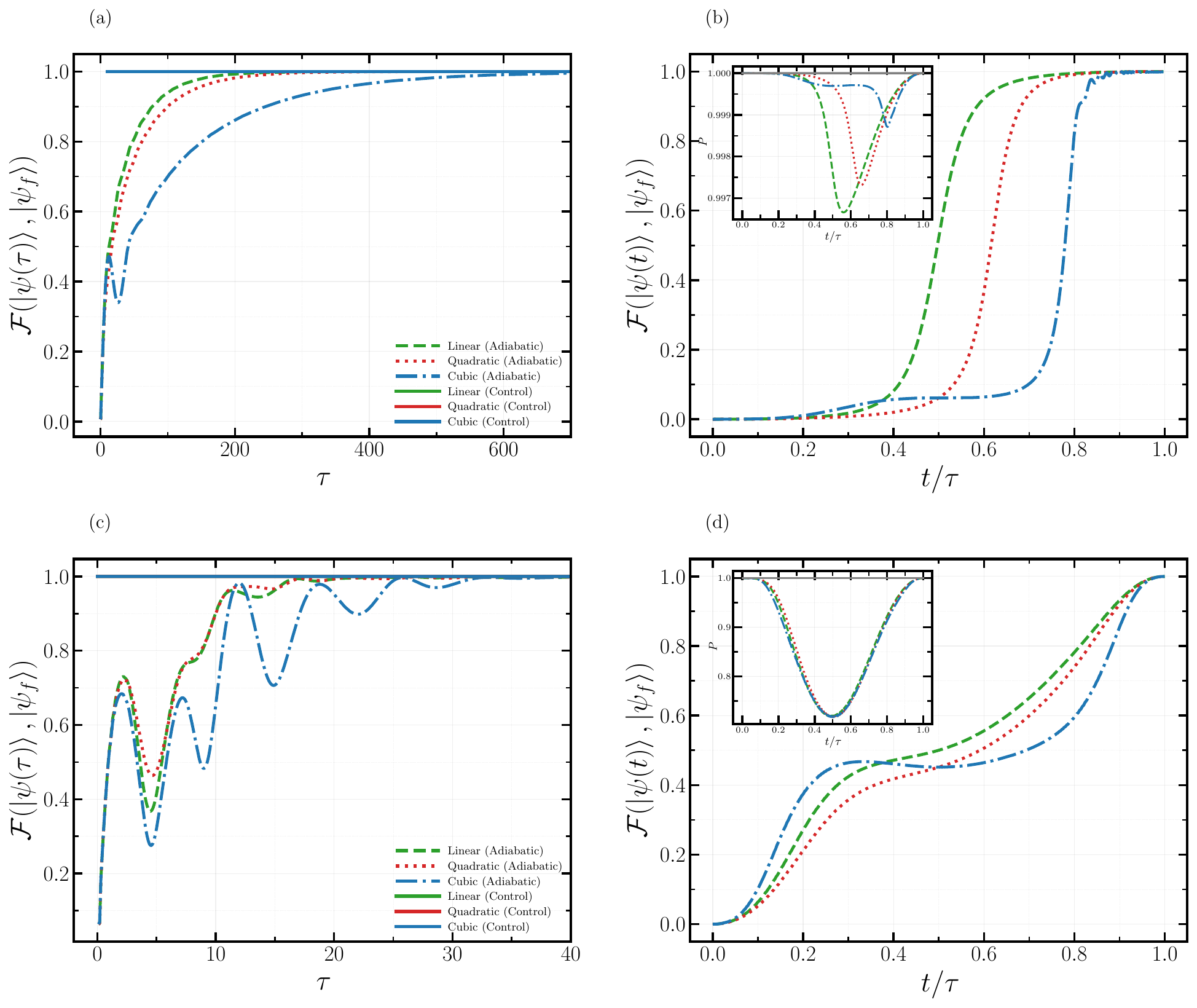}
    \caption{The target state $\mathcal{F}(\ket{\psi(\tau)}, \ket{\psi_i})\!=\! |\bra{\psi(\tau)}\ket{\psi_f}|^2$ and 
    the instantaneous fidelity $\mathcal{F}(\ket{\psi(t)}, \ket{\psi_f}) = |\bra{\psi(t)}\ket{\psi_f}|^2$ is plotted as a function of $\tau$ and $t/\tau$ for $\tau\!=\!1000$ respectively for different driving profiles. Panel (a) and (b) [(c) and (d)] show the weak-coupling [strong-coupling] regime of Rabi model for transfer of initial state $\ket{\psi_i}=\ket{e,0}$ to the target state $\ket{\psi_f}\!=\!\ket{g,1}$ with and without control. The green-dashed curves corresponds to the linear driving, the red-dotted curves represent the quadratic driving, and the blue-dashed-dotted curve is the cubic driving. 
    Parameters used are $\omega_i\!=\!-\omega_c/2$, $\omega_f\!=\!5\omega_c/2$, $\lambda_0\!=\!0$, $\alpha\!=\!0.5$, and $\beta\!=\!2$. For the coupling protocol; $\lambda_m\!=\!\omega_c/10$ (panel (a) and (b)) for weak-coupling limit, and  $\lambda_m\!=\!\omega_c$ (panel (c) and (d)) for strong-coupling limit.}  
\label{fig:obinna fidelity}
\end{figure} 
In Fig.~\ref{fig:obinna fidelity}(a) and (b), we plot the  fidelity of the target state, $\mathcal{F}(\ket{\psi(\tau)}, \ket{\psi_i})\!=\! |\bra{\psi(\tau)}\ket{\psi_f}|^2$, as a function of $\tau$, and the instantaneous fidelity $\mathcal{F}(\ket{\psi(t)}, \ket{\psi_f}) = |\bra{\psi(t)}\ket{\psi_f}|^2$ as a function of  $t/\tau$ for $\tau\!=\!1000$, where $\ket{\psi_f} = \ket{g,1}$ and $\ket{\psi(t)} = \hat{U}_{RM}(t)\ket{e,0}$. It can be seen that increasing the complexity of the driving increases the amount of time required to achieve complete population inversion of the state, i.e., $\mathcal{F}\simeq 1$. However, as expected, the transfer of the perfect state always occurs in the adiabatic limit, $\tau>600$, irrespective of the driving protocol. Yet, with control, we achieve a perfect population inversion for any driving time $\tau$ and for any form of driving. The inset in Fig.~\ref{fig:obinna fidelity}(b) shows the occupation probability of the state being found in the subspace $\mathcal{H}_0$ during the adiabatic driving computed with $ P_j(t)\!=\!\mathrm{Tr}[P_j\rho(t)]$, where $\rho(t) = \ket{\psi(t)}\bra{\psi(t)}$, $P_j\!=\!\sum_{\ket{\phi} \in \mathcal{H}_j} \ket{\phi}\bra{\phi}$ and $j$ labels the subspace. As can be seen, in the weak coupling regime, the state remains within the zeroth ($n\!=\!0$) subspace, in  good agreement to the RWA prediction.

\begin{figure}[t] 
\hspace*{-0.5cm}
    \includegraphics[width = 0.5\textwidth]{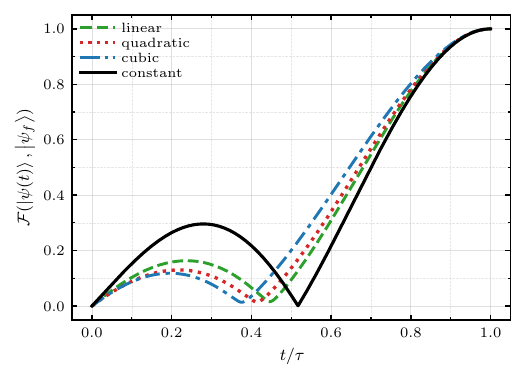}
    \caption{The instantaneous fidelity $\mathcal{F}(t)\!=\!|\langle\psi_f|\psi(t)\rangle|^2$ between the evolved state $\ket{\psi(t)}$ and the target states $\ket{\psi_f}=\frac{1}{\sqrt{2}}(\ket{g,1}+\ket{e,5})$ as a function of time for $\tau\!=\!1$. The system is initialised in $\ket{\psi_i}=\ket{e,0}$ and evolved under $\hat{H}(t)=\hat{H}_{RM}(t)+\hat{H}_c(t)$ using the driving protocols, see Eq.~(\ref{coupling protocol}) and Eq.~(\ref{TLS driving}). The green-dashed curves corresponds to the linear driving, the red-dotted curves represent the quadratic driving, and the blue-dashed-dotted curve is the cubic driving. The parameter used are $\omega_i=-\omega_c/2$, $\omega_f=5\omega_c/2$, $\alpha\!=\!0.5$, $\beta\!=\!2$,   $\lambda_0=0$ and $\lambda_m=\omega_c/10$. The black solid curve depicts the constant qubit frequency (no driving) case, where we  used  $\omega_i=\omega_f=5\omega_c/2$ and $\lambda_0=\lambda_m=0$. 
}
\label{fig:combined_single_shot_fidelity}
\end{figure} 


To generate entangled cat state using the adiabatic protocol, a $\pi/2$-pulse is applied to the prepared Fock state $\ket{e, n}$  to realize the superposition $(\ket{e, n}+\ket{g, n})/\sqrt{2}$. After the population inversion, an entangled cat state of the form $\ket{\psi_{n+1,n-1}}\!=\!(\ket{g,n+1}+\ket{e,n-1})/\sqrt{2}$ is obtained, which can then be transformed into $\ket{\psi_{n+2,n-2}}\!=\!(\ket{g,n+2}+\ket{e,n-2})/\sqrt{2}$ by applying another $\pi$ pulse and population inversion. Repeating this procedure allows us to build any entangled cat state of the form $\ket{\psi_{n+k,n-k}}\!=\!(\ket{g,n+k}+\ket{e,n-k})/\sqrt{2}$ where $1\!\leq\!k\!\leq\!n$. 
The generation of such cat state can be sped up using the time-optimal (quantum wind) control for any given $\tau$. Additionally, since the wind control can drive any initial state $\ket{\psi_i}$ to the target state $\ket{\psi_f}$, there is no need for the  sequential generation of the Fock state $\ket{e,0} \rightarrow\ket{g,1} \rightarrow \ket{e,1} \rightarrow \ket{g,2} \cdots$. This is apparent in Figure \ref{fig:combined_single_shot_fidelity} which presents the  instantaneous fidelity between the target state $\ket{\psi_f}\!=\!(\ket{g,1}+\ket{e,5})/\sqrt{2}$ and $\ket{\psi(t)}$ obtained via evolving  the initial state $\ket{e,0}$ under Eq.~(\ref{RM model0}) and the quantum wind control, i.e., total Hamiltonian $\hat{H}(t)\!=\!\hat{H}_{RM}(t)+\hat{H}_c(t)$. It can be seen  there is a perfect transfer of the quantum state to the target state which can be achieved for any given $\tau$, when evolved with the total Hamiltonian $\hat{H}(t)$. 
Note, we selected $\ket{\psi_i} = \ket{e,0}$ and $\ket{\psi_f}=(\ket{g,1}+\ket{e,5})/\sqrt{2}$ since it can be prepared using the sequential adiabatic population inversion scheme previously discussed. However, the quantum wind control is not restricted by such constraints and, in principle, one can generate any arbitrary states within the Hilbert space $\mathcal{H}$. To illustrate, we also consider the transition $\ket{\psi_i} = \ket{g,0} \rightarrow \ket{\psi_f}\!=\!(\ket{g,0}+\ket{e,5})/\sqrt{2}$ which we generate using the quantum wind control (results not shown), despite being inaccessible through adiabatic evolution.

Therefore, the quantum wind protocol has clear advantages over STA protocols such as counteradiabatic driving \cite{PhysRevLett.124.180401} and local counterdiabatic driving \cite{PhysRevLett.111.100502} for generation of nonclassical state. That is, it  offers the possibility of  a single shot protocol, it can generate non-adiabatic solutions and there no restrictions on the type of driving protocol used in the state generation. An interesting example of this is for time independent Hamiltonian, $\hat{H}_0\!=\!\omega_c\hat{a}^\dagger \hat{a} + \frac{\omega_f}{2}\hat{\sigma}_z$, where we used $f(\mathcal{S})\!=\!1$, and $\lambda(\mathcal{S})\!=\!0,  \forall \mathcal{S}$. The control Hamiltonian that drives the initial state $\ket
{\psi_i} = \ket{g,0}$ towards the target states $\ket{\psi_f}\!=\!(\ket{g,0}+\ket{e,n})/\sqrt{2}$ then reads, in the Schr\"{o}dinger picture,
$\hat{H}_{c}^{(S)}(t)
=
\frac{i\pi}{4\tau}
\left(
e^{-i(n\omega_{c}+\omega_{f})t}
e^{in\omega_{c}\tau}
\ket{e,n}\bra{g,0}
-
\mathrm{h.c.}
\right)$. 


\emph{Strong-coupling regime - Beyond RWA} -- For $\eta = \lambda/\omega_c>0.1$, the RWA is no longer valid and the system cannot be described by a two-level-like Hamiltonian.
In this regime, the Rabi model $\hat{H}_{RM}$ in Eq.~(\ref{RM model0}) conserves the total number of qubit and bosonic excitations modulo 2, i.e., the parity of the total excitation  described by the operator $\hat{P} = \hat{\sigma}_z\hat{\Pi}$ is conserved $[\hat{P}, \hat{H}_{RM}]=0$ where $\hat{\Pi}=(-1)^{\hat{a^\dagger \hat{a}}}$ is the parity operator of the cavity. Moving via unitary transformation to the frame where the conservation is explicit results in the parity-frame Hamiltonian $\hat{H}_{\pm}$ \cite{leroux2017simple}. The parity-frame Hamiltonian $\hat{H}_{\pm}$ can be written as a direct sum of the Hamiltonians $\hat{H}_-$ and $\hat{H}_+$, which act on the subspace defined by $\langle \sigma_z\rangle=\pm 1$. That is,  $\hat{H} = \hat{H}_+ \oplus \hat{H}_-$, where $\mathcal{H}_{+}\!=\!\mathrm{span} \left\{
\ket{e,0},
\ket{g,1},
\ket{e,2},
\ket{g,3},
\ldots
\right\},$ and $\mathcal{H}_{-} \!=\!\mathrm{span}
\left\{
\ket{g,0},
\ket{e,1},
\ket{g,2},
\ket{e,3},
\ldots
\right\}.$
Despite the fact higher order Fock states becoming populated during the driving $\ket{\psi(t)} \in \mathcal{H}_+$, we find that population inversion between $\ket{e,n}$ and $\ket{g,n+1}$ is still possible using the same sequential adiabatic protocol previously used assuming the coupling vanishes at the boundaries, i.e., $\lambda(0)\!=\!\lambda(\tau)\!=\!0$.  This is shown in Fig.~\ref{fig:obinna fidelity}(c) and (d), where we present the target state fidelity, and the instantaneous fidelity as a function of $\tau$ and $t/\tau$, respectively. We find that there is a perfect state transfer in a shorter time, i.e. $\tau\simeq 40$, compared to the weak-coupling limit. Linear and quadratic driving protocols have similar behaviour, whereas the cubic driving protocol results in fluctuations in the small time limit. 
 
Although, Fock state $\ket{e,n}$  can still be generated adiabatically without the RWA,  numerical analysis show that it is not possible to obtain entangled cat states of the form $\ket{\psi_{N+k,N-k}}\!=\!(\ket{g,N+k}+\ket{e,N-k})/\sqrt{2}$. This because perfect adiabatic population inversion of $(\ket{e,N}+\ket{e,N})/\sqrt{2}$ is only possible in the weak coupling regime, since the state stays approximately within $\mathcal{H}_0$, see the insert in Fig.~\ref{fig:obinna fidelity}(b). However, in contrast, beyond the RWA the state lies with the $\mathcal{H}_+$ subspace, see the insert in Fig.~\ref{fig:obinna fidelity}(d). Therefore, in the strong coupling regime one must use the quantum wind to generate $\ket{\psi_{N+k,N-k}}$, as demonstrated for the weak-coupling limit in Fig.~\ref{fig:combined_single_shot_fidelity}.



\emph{Deep strong coupling (DSC) regime} -- In this regime, when $\lambda \gg \omega_q$ and $\eta\gtrsim 1$, the Rabi model Hamiltonian   can be approximated  as $\hat{H}_{RM}(t) \approx \omega_c\hat{a}^{\dagger}\hat{a} + \lambda(t)\hat{\sigma}_x (\hat{a}^\dagger +\hat{a})$. 
Therefore, in the DSC regime the ground state of $\hat{H}_{RM}$ is the so-called giant entangled cat state \cite{leroux2017simple}, written as 
\begin{equation} \label{Giant entangled cat state}
    \ket{G} = \frac{1}{2}(\mathcal{N}_{+}\ket{g, \text{Cat}_{+}}-\mathcal{N}_{-}\ket{e,\text{Cat}_{-}})
\end{equation}
where $\ket{\text{Cat}_{\pm}}\!=\!\frac{1}{\mathcal{N}_{\pm}}(\ket{\eta} \pm e^{i \theta}\ket{-\eta}$ is an optical cat state, $\mathcal{N_{\pm}}\!=\!\sqrt{2(1 \pm e^{-2|\eta|^2)}}$ and $\theta$ is some phase between the coherent states. 
To understand this, the last term of the Rabi model, Eq.~(\ref{RM model0}), $V=\omega_q(t)\hat{\sigma}_z/2$ can be treated as a small perturbation. For $\omega_q(t) \ll1$, $\hat{\sigma}_x$ becomes approximately conserved $[\hat{\sigma}_x ,\hat{H}_{RM}(t)] = i \omega_q(t)\hat{\sigma}_y \approx 0$
and the dominant contribution becomes \cite{Li2021}
\begin{equation}
\hat{H}_0 \equiv \hat{H}_{RM} (\omega_q=0) = \omega_c \hat{a}^{\dagger}\hat{a}
+ \lambda(t)\hat{\sigma}_x(\hat{a}^{\dagger}+\hat{a}),
\end{equation}
which may be diagonalised using the polaron transformation;
\begin{equation}
\hat{U}_{P}
= |+x\rangle\langle +x|\,\hat{D}(\eta)
+ |-x\rangle\langle -x|\,\hat{D}(-\eta),
\end{equation}
where $\hat{D}(\eta)=\exp(\eta \hat{a}^{\dagger}-\eta^{*}\hat{a})$
is the displacement operator and $\eta(t)=-\lambda(t)/\omega_c$ and $\hat{\sigma}_x \ket{\pm} = \pm \ket{\pm}$. Hence, the eigenstates of $\hat{H}_0$ are then given by the displaced Fock states
\begin{equation}
|\psi_{\pm,n}\rangle
= \hat{D}(\pm\eta)|n\rangle |\pm x\rangle,
\end{equation}
which are degenerate with respect to the qubit state, restricting the system to the degenerate ground-state manifold spanned $\{|\eta\rangle|+x\rangle,  |-\eta\rangle|-x\rangle\}$.
However, the perturbation term couples these degenerate states, and the resulting eigenstates become the symmetric and antisymmetric parity superpositions
\begin{equation}
|\psi_{\pm}^{\mathrm{ECS}}\rangle
= \frac{1}{\sqrt{2}}
\left(
|\eta\rangle|+x\rangle \pm |-\eta\rangle|-x\rangle
\right).
\end{equation}
the expanding $\ket{\pm x}\!=\!(\ket{g} \pm \ket{e})/\sqrt{2}$ gives the giant entangled cat state, Eq.~(\ref{Giant entangled cat state}).


Therefore, adiabatically evolving the initial ground state $\ket{g,0}$ of Rabi Hamiltonian $\hat{H}_{RM}(t)$ in the deep strong-coupling limit can generate entangled cat states $\ket{G}$, provided at the end of the driving $\eta(\tau)\sim 1$ and $\lambda(\tau)\gg\omega_q(\tau)$. This is achieved using the strong coupling protocol given in Eq.~\ref{strong coupling protocol} 
\begin{equation} \label{strong coupling protocol}
    \lambda(t) = (\lambda_m -\lambda_0)\sin\left(\frac{\mathcal{S}\pi}{2} + \phi\right) + \lambda_0,
\end{equation}
such that $\lambda(\tau) \gg \omega_q(\tau)$ is satisfied. 


\begin{figure}[!h] 
\hspace*{-0.5cm}
    \includegraphics[width = 0.5\textwidth]{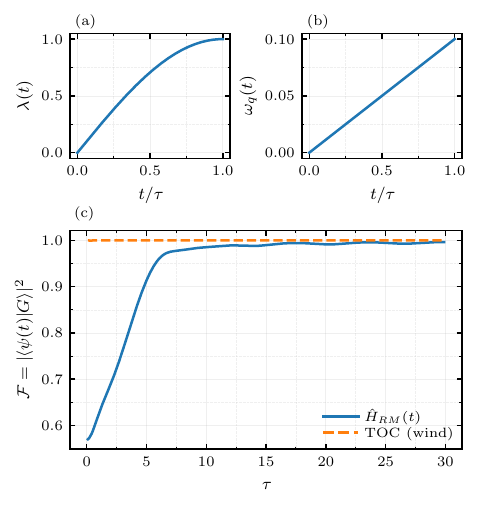}
    \caption{(a) The time-dependent coupling protocol, Eq.~(\ref{strong coupling protocol}), (b) the qubit driving frequency (linear case, $j=1$), Eq.~(\ref{TLS profiles}), and (c) the target state fidelity $\mathcal{F}(\tau)=|\langle\psi_f|\psi(\tau)\rangle|^2$ as a function of time. We used parameters $\omega_i=0$, $\omega_f=\omega_c/10$, $\lambda_0=0$, and $\lambda_m=\omega_c$ to obtain $\ket{\psi(\tau)}$ by evolving $\ket{\psi_i} = \ket{g,0}$ towards $\ket{G}\!=\!\frac{1}{2}(\mathcal{N}_{+}\ket{g, \mathrm{Cat}_{+}}-\mathcal{N}_{-}\ket{e,\mathrm{Cat}_{-}})$ with (orange) and without (blue) control.}
\label{fig:Sam fidelity}
\end{figure} 
In this framework, we can directly compare the adiabatic regime to the quantum wind control used as a shortcut-to-adiabatic driving for generating highly non-classical entangled cat states (since both are single shot protocols). In Fig.~\ref{fig:Sam fidelity}(a) and Fig.~\ref{fig:Sam fidelity}(b) we present the driving functions we will be using in the DSC regime. In Fig.~\ref{fig:Sam fidelity}(c), we present the fidelity $\mathcal{F}(\tau)=|\langle\psi_f|\psi(\tau)\rangle|^2$ between the target state $\ket{\psi_f}\!=\!\frac{1}{2}(\mathcal{N}_{+}\ket{g, Cat_{+}}-\mathcal{N}_{-}\ket{e,Cat_{-}})$ and $\ket{\psi(\tau)}$  as a function of the total driving time $\tau$.  The state $\ket{\psi(\tau)}$ is obtained by evolving the initialized state $\ket{\psi_i}=\ket{g,0}$ without control (blue) $\hat{H}(t) = \hat{H}_{RM}(t)$ and with control (orange) $\hat{H}(t) = \hat{H}_{RM}(t) + \hat{H}_c(t)$. It can be seen that the generation of an entangled cat state can be generated in an arbitrarily short time using the wind control.




\subsection{Non-classicality of the entangled cat state}\label{nonclassicality}
The nonclassical features of quantum states are commonly characterized through the Wigner quasiprobability distribution \cite{PhysRev.40.749}. However, this formulation is generally restricted to continuous-variable systems, such as cavity fields. For hybrid systems consisting of both discrete and continuous degrees of freedom, such as qubit-cavity states, a complete phase-space representation can instead be constructed using the set of joint Wigner functions, see \cite{vlastakis2015characterizing}, defined as
\begin{equation}
\label{joint Wigner function}
W_i(\beta,\rho)
=\frac{2}{\pi}
\expval{\hat{\sigma}_i P_\beta}_{\rho},
\end{equation}
where 
$P_{\beta}
=
D(\beta)
\hat{\Pi}
D^{\dagger}(\beta)$
is the displaced parity operator,
$D(\beta)
=
\exp{\beta\hat{a}^{\dagger}-\beta^*\hat{a}}$
is the displacement operator, and 
\(
i=\{I,X,Y,Z\}
\)
labels the identity and Pauli operators. The hybrid qubit-cavity density matrix may then be reconstructed as
$\rho(t)=\pi\sum_i\int W_i(\beta,\rho(t))\,\sigma_i P_\beta\, d^2\beta.$
The set of joint Wigner functions therefore provides a complete representation of the hybrid qubit-oscillator state, while simultaneously revealing nonclassical features of the conditional oscillator states and correlations between the qubit and bosonic degrees of freedom.

\begin{figure*}[thp!]
    \centering
    \includegraphics[width = 0.9\textwidth]{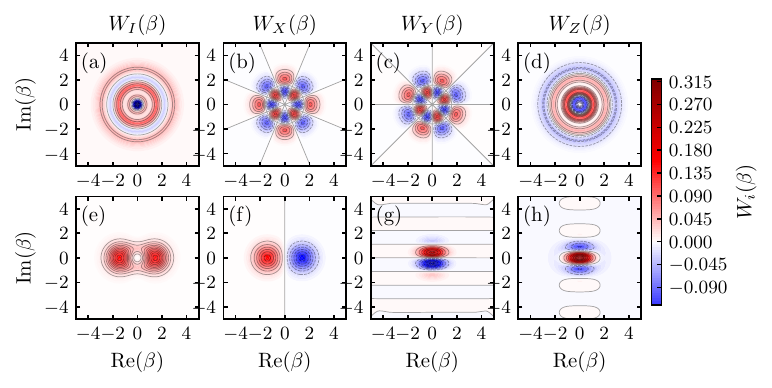}
    \caption{The set of joint Wigner function $W_i(\beta, \rho(\tau))$
    of the engineered target states $\ket{\psi_{5,1}}\!=\!\frac{1}{\sqrt{2}}(\ket{g,1}+\ket{e,5})$ and $\ket{G}\!=\!\frac{1}{2}(\mathcal{N}_{+}\ket{g, \mathrm{Cat}_{+}}-\mathcal{N}_{-}\ket{e,\mathrm{Cat}_{-}})$ is plotted in (a)-(d) and (e)-(h) respectively for $i =\{I, X, Y, Z \}$. The engineered states $\rho(\tau) = \ket{\psi(\tau)}\bra{\psi(\tau)}$ are obtained via evolving the initialised state $\ket{\psi_i} = \ket{g,0}$ under $\hat{H}(t)=\hat{H}_{RM}(t)+\hat{H}_c(t)$ at $\tau=1$. In (a)-(b) we used the driving protocols given by 
    Eq.~(\ref{coupling protocol}) and Eq.~(\ref{TLS profiles}) (linear) with $\omega_i=-\omega_c/2$, $\omega_f\!=\!5\omega_c/2$, $\lambda_0\!=\!0$ and $\lambda_m\!=\!\omega_c/10$. Parameters used in (c)-(d), Eq.~(\ref{TLS profiles}) (linear) with $\omega_i=-\omega_c/2$, $\omega_f\!=\!5\omega_c/2$, and $\lambda(t)$ according to Eq.~(\ref{strong coupling protocol}) with $\lambda_0=0$ and $\lambda_m=\omega_c$.}
\label{fig:negativity of final states}
\end{figure*}

Figure~\ref{fig:negativity of final states} displays the joint Wigner functions of the engineered entangled cat states in both the strong $\ket{G}$- and weak-coupling $\ket{\psi_{n+k,n-k}}$ regimes for $\tau=1$. In the presence of control, the target states are prepared with unit fidelity in both regimes. In contrast, without control the achievable fidelity is significantly reduced, reaching only $\mathcal{F}(\tau)\approx0.624$ in the deep-strong-coupling regime and $\mathcal{F}(\tau)\approx0.087$ in the weak-coupling regime.

\subsection{Trade-off between speed and energetic  cost}\label{sec.cost}
\begin{figure}[h!] 
\hspace*{-0.5cm} 
    \includegraphics[width = 0.5\textwidth]{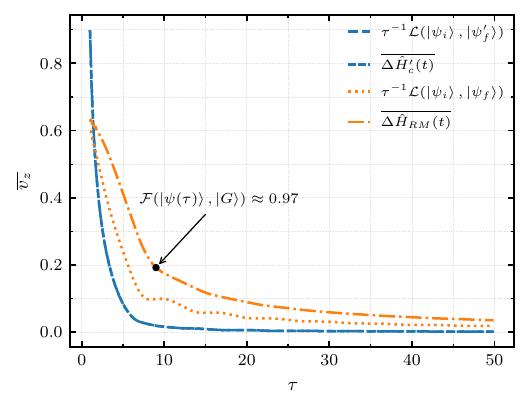}
    \caption{The energy resource $\overline{v_z}$ and its bound is plotted as a function of driving time $\tau$ using $\overline{v_z} = \overline{\Delta \hat{H}(t)}$ and $\overline{v_z} = \tau^{-1}\mathcal{L}(\ket{\psi_i, \ket{\psi_f}})$ respectively where $\ket{\psi_i} = \ket{g,0}$ and $\ket{\psi_f}$ is obtained by evolving the initial state with (blue) and without (orange) control. Parameters used $\omega_i=0$, $\omega_f=\omega_c/10$, $\lambda_0=0$, and $\lambda_m=\omega_c$}
\label{fig:speed limit}
\end{figure}

Quantum theory fundamentally limits the speed of evolution of a system between given initial and final states, see \cite{Deffner2017} for review. Specifically, the minimum time taken for a quantum system to evolve from an initial state to a final state is called the quantum speed limit (QSL) time. 
Mandelstam and Tamm derived the first expression of the quantum QSL time, which relates the minimum evolution time to both
the geometric distance between quantum states in projective Hilbert space and the energetic resources available during the dynamics \cite{mandelstam1945uncertainty,Mandelstam1991} i.e,
\begin{equation}
\label{MT bound}
\tau \geq \tau_{MT}
= \frac{\mathcal{L}(\ket{\psi_i},\ket{\psi_f})}{\overline{\Delta \hat{H}(t)}},
\end{equation} 
where $\tau_{MT}$ is the QSL time, $\ket{\psi_f} = \hat{U}(\tau)\ket{\psi_i}$ and 
$\overline{\Delta \hat{H}(t)}\!=\!\tau^{-1}\int_0^\tau \Delta \hat{H}(t)dt$ is the time-averaged energy variance of $\hat{H}(t)$/energetic resource. For perfect time optimal evolution, no energetic resources are wasted and the MT bound saturates such that
\begin{equation} \label{minima}
    \overline{\Delta \hat{H}(t)} = \frac{\mathcal{L}(\ket{\psi_i},\ket{\psi_f})}{\tau}
\end{equation}
 Within the quantum wind framework the equality condition is satisfied by design such that $\tau\!=\! (\Delta \hat{H}_c')^{-1} \mathcal{L}(\ket{\psi_i},\ket*{\psi_f'})$.
Hence, implementing the wind control can be done so with minimal energetic waste. We present this in Figure~\ref{fig:speed limit} where the energetic resource is plotted as a function of $\tau$ with and without control. 
 \newline 
It shows that the wind control saturates the bound as expected and therefore is completely energy resource efficient. Whereas in the adiabatic regime the bound does not saturate and furthermore high-fidelity state preparation requires substantially longer evolution times and larger energetic resources, as shown in Fig.~\ref{fig:Sam fidelity}. 
Additionally, this bound can be explicitly related to the cost of the implementation of the control using the fact the quantum "speed" $v_z(t) = \Delta \hat
H(t)$ is bounded by Eq.~\ref{cost bound}
\begin{equation} \label{cost bound}
\Delta \hat H(t)
=
\sqrt{
\langle \hat H^2(t)\rangle
-
\langle \hat H(t)\rangle^2
}
\leq
\|\hat H(t)\|,
\end{equation}
where $||*||$ denotes the Hilbert--Schmidt norm, which is commonly used to quantify control costs \cite{Zheng2016,abah2019energetic} via $C=\overline{\|\hat H(t)\|} = \tau^{-1}\int_0^\tau|\hat H(t)\|dt$. Therefore, one can obtain 
\begin{equation}
    C \geq \frac{\mathcal{L}(\ket{\psi_i},\ket{\psi_f})}{\tau}
\end{equation}
For the wind a special case emerges where $\hat
H'_c$ is traceless and therefore $\Delta \hat
H'_c = \|\hat H'_c\|$ \cite{garcia2022highly}. Hence, there is a clear trade off between the speed at which one can generate the non-classical states and the energetic cost required to do so. However, for "favourable winds" the lower bound of the cost of implementable can be decreased due to the geometry of the rotated frame i.e $\mathcal{L}(\ket{\psi_i},\ket*{\psi'_f}) < \mathcal{L}(\ket{\psi_i},\ket{\psi_f})$. Hence, the wind control may be most practical for the generation of states for very small $\tau$. \newline
\newline

\section{Robustness of the protocol} \label{sec:Noise}
Here, we focus on discussing the robustness of the quantum wind protocol.
For practical preparation of the quantum state, the control protocol must remain reliable in the presence of experimental imperfections and environmental decoherence. Since quantum wind control is derived independently of the underlying dynamics generated by the original Hamiltonian $\hat{H}_0(t)$, it is intrinsically robust against inaccuracies in the implementation of driving fields $\omega_q(t)$ and $\lambda(t)$. However, the wind protocol is sensitive to interactions with the environmental. To quantify the effects of cavity and qubit losses, we model the open-system dynamics using the Bloch--Redfield master equation \cite{bloch1957generalized} with partial secularization
\begin{align}
\label{Redfield}
\frac{d}{dt}\rho_{mn}(t)
&=
-i\omega_{mn}(t)\rho_{mn}(t)
+
\sum_{j,k}
\Big[
R_{mjjk}\rho_{kn}
+
R_{nkkj}^{*}\rho_{mj}
\notag\\
&\quad
-
R_{knmj}\rho_{jk}
-
R_{jnmk}^{*}\rho_{jk}
\Big],
\end{align}
where $\omega_{mn}(t)=E_m(t)-E_n(t)$ are the instantaneous Bohr frequencies and the Redfield tensor is determined by the system-bath coupling operators in the instantaneous energy basis. We neglect the Lamb-shift contribution and consider only dissipative processes characterized by thermal transition rates governed by Bose--Einstein statistics.
\begin{figure}[t!] 
    \includegraphics[width = 0.5\textwidth]{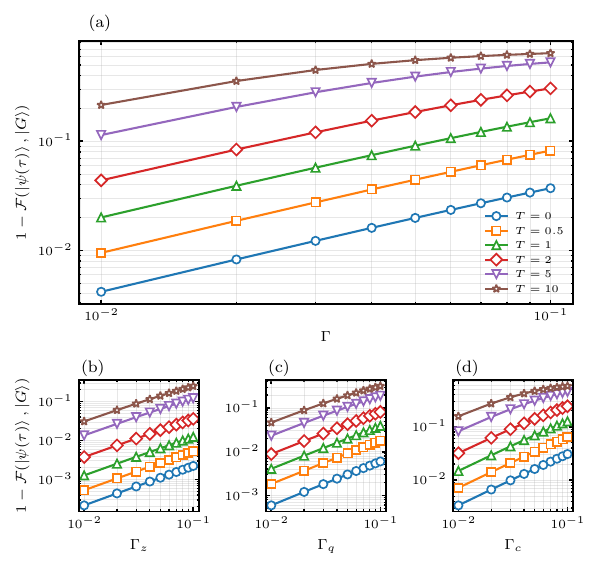}
    \caption{The infidelity of the target state $1-\mathcal{F}(\ket{\psi(\tau)}, \ket{G})$ where $\ket{G} = \frac{1}{2}(\mathcal{N}_{+}\ket{g, Cat_{+}}-\mathcal{N}_{-}\ket{e,Cat_{-}})$ and $\ket{\psi(\tau)}$  as a function of cavity/qubit loss rates for different temperature $T$. We assume that the initial state is  $\ket{\psi_i}\!=\! \ket{g,0}$ and $\tau=1$.  We used the coupling driving protocol $\lambda(t)$ in Eq.~(\ref{strong coupling protocol}) and the linear form of $\omega_q(t)$,  Eq.~(\ref{TLS profiles}), where $\omega_i\!=\!\omega_f = 0.1\, \omega_c$, $\lambda_0=0$ and $\lambda_m=\omega_c$.  Parameter used: (a) $\Gamma\!=\!\Gamma_z\!=\!\Gamma_q\!=\!\Gamma_c$, (b) $ \Gamma_q\!=\!\Gamma_c\!=\!0$, (c) $\Gamma_z\!=\!\Gamma_c\!=\!0$, and (d) $\Gamma_q\!=\!\Gamma_c\!=\!0$}.
\label{fig:decoherence}
\end{figure} 
To account for realistic noise mechanisms, we consider system-bath coupling operators
$V_\alpha=
\left\{
\sqrt{\Gamma_c}(\hat{a}+\hat{a}^{\dagger}),
\sqrt{\Gamma_q}\hat{\sigma}_x,
\sqrt{\Gamma_\phi}\hat{\sigma}_z
\right\}$,
which describe cavity dissipation, qubit relaxation/excitation, and qubit dephasing, respectively. 
Figure~\ref{fig:decoherence} shows the target-state infidelity,
$
1-\mathcal{F}(\ket{\psi(\tau)},\ket{\psi_f}),
$
for the target state $\ket{\psi_f}=\ket{G}$ with $\tau=1$. The results demonstrate that the quantum wind protocol maintains high-fidelity state preparation across a broad range of environmental conditions and temperatures. The robustness originates from the reduced evolution time, which minimizes the system exposure to dissipative processes. Similarly to shortcut-to-adiabaticity protocols \cite{abah2020quantum,chen2021shortcuts,liu2022generation}, accelerated preparation significantly outperforms adiabatic evolution in the presence of cavity and qubit losses, where long evolution times lead to substantial degradation of the target-state fidelity.

\section{Conclusions} \label{sec:Conculsion}
Using time time-optimal control techniques namely the so called quantum wind we have presented a framework for the fast generation of nonclassical states in light--matter systems described by the Jaynes--Cummings and quantum Rabi models. 
We demonstrated the preparation of Fock states and highly entangled Schr\"odinger cat states in both weak- and strong-coupling regimes. The nonclassical features of the engineered states have been characterized using the joint Wigner functions in order to display the non-classicality of both correlations and the cavity itself. To understand the fundamental performance limits of the protocol, we analyze the associated quantum speed limits and their relation to energetic resources. We showed that the protocol operates at the minimum-time bound while reducing the energetic cost required for state preparation through favourable winds. Furthermore, our analysis revealed strong robustness against cavity losses, qubit relaxation, dephasing, and finite-temperature effects. The enhanced performance originates from the shortened preparation times, which suppress exposure to environmental noise and dissipative processes.
Our results demonstrate that time-optimal control provides an efficient and robust route for generating highly nonclassical states and may serve as a promising tool for quantum state engineering in state-of-the-art cavity and circuit quantum electrodynamics platforms as well as other hybrid quantum architectures.


\bibliography{bibliography} 
\end{document}